\documentclass[prl,superscriptaddress,twocolumn,preprintnumbers]{revtex4}

\usepackage{color}
\usepackage[active]{srcltx}
\usepackage{amsmath,amsfonts,amssymb,amsthm,amstext,amscd,eucal,srcltx}
\usepackage{epsfig,graphicx,bm}
\usepackage{epstopdf, epsf}
\usepackage{dcolumn}
\usepackage{hyperref}

\newcommand{\be}{\begin{equation}}
\newcommand{\ee}{\end{equation}}

\newcommand{\bse}{\begin{subequations}}
\newcommand{\ese}{\end{subequations}}
\newcommand{\bea}{\begin{eqnarray}}
\newcommand{\eea}{\end{eqnarray}}
\newcommand{\ba}{\begin{array}}
\newcommand{\ea}{\end{array}}
\newcommand{\bc}{\begin{center}}
\newcommand{\ec}{\end{center}}

\begin{document}
\preprint{IPM/P-2012/009}  
\vspace*{3mm}
\title {A Unified picture of Dark Matter and Dark Energy from Invisible QCD}%

\author{Andrea Addazi}
\email{andrea.addazi@lngs.infn.it}
\affiliation{Dipartimento di Fisica,
Universit\`a di L'Aquila, 67010 Coppito AQ and LNGS, Laboratori Nazionali del Gran Sasso, 67010 Assergi AQ, Italy}

\author{Stephon Alexander}
\email{Stephon\_Alexander@brown.edu}
\affiliation{Department of Physics, Brown University, Providence, Rhode Island 02912, USA}

\author{Antonino Marcian\`o}
\email{marciano@fudan.edu.cn}
\affiliation{Center for Field Theory and Particle Physics \& Department of Physics, Fudan University, 200433 Shanghai, China}

\begin{abstract}

\noindent 
It has been shown in a companion paper that the late time acceleration of the universe can be accounted for by an extension of the QCD color to a $SU(3)$ invisible sector (IQCD).  In this work we discuss a unified framework such the scale of dark chiral-breaking dictates both the accelerated expansion of the universe, and the origin of dark matter. We find that the strong and gravitational dynamics of dark quarks and gluons evolve to eventually form exotic dark stars.  We discuss the dynamical complexity of these dark compact objects in light of dark big bang nucleosynthesis. We argue how IQCD favors a halo composed of very compact dark neutron stars, strange/quark stars and black holes, with masses $M_{MACHO}< 10^{-7}M_{\odot}$. This avoids limit from MACHO and EROS collaborations as well as limit from clusters. 
We also discuss possible phenomenological implications in dark matter searches. 
We argue that dark supernovae and dark binaries can emit very peculiar gravitational waves signal 
testable by the LIGO/VIRGO collaboration and future projects dedicated to these aspects. 

 \end{abstract}
\pacs{98.80.Cq}
\keywords{Dark Energy, non-Abelian gauge theory, Condensate}
\maketitle

\section{Introduction}

\noindent

While there is experimental evidence that the universe is accelerating \cite{Riess:1998cb,Perlmutter:1998np,Frieman:2008sn, Astier:2012ba, Spergel:2003cb, Graham:2005xx, Spergel:2006hy, Komatsu:2008hk,Cole:2005sx,Tegmark:2006az}, compelling theoretical explanations and predictions are still under investigation (see {\it e.g.} \cite{Capozziello:2003tk, Carroll:2003wy, AT}).  It is still possible that a pure cosmological constant can explain the acceleration, but because of the unnatural fine tuning, it is possible that Dark Energy might be due to a dark dynamical field (for a recent attempt see also \cite{Dona:2015xia}), where the phenomenology of these new fields may naturally be connected to another unresolved problem-dark matter (see {\it e.g.} \cite{Addazi:2016sot}).

In this work, we show that the dynamical model of dark energy, presented in \cite{IQCD}, can naturally predict dark matter (DM).  We will pursue this possibility by revisiting constraints on the coupling to the visible sector so as to be simultaneously consistent with dark-matter and dark energy.  The idea of an invisible QCD sector is strongly motivated by several different UV completions of the standard model and quantum gravity. For example $E_{8}\times E_{8}$ heterotic superstring theory inevitably leads to the presence of dark gauge sectors. In IIA and IIB open string theories, the Standard Model can be completed by intersecting D-branes wrapping n-cycles in the internal $CY_{3}$\footnote{Also in these models, the presence of dark sectors is usually requested in order to cancel stringy tadpole anomalies and to introduce discrete symmetries stabilizing the proton \cite{Anastasopoulos:2015bxa}. We also mention that non-perturbative stringy effects called exotic instantons can lead to interesting implications in particle physics and baryogenesis, as the generation of a $\Delta B=2$ effective Majorana mass term for the neutron \cite{Addazi:2014ila} Alternatively, a dark copy is naturally obtained extending the spin connection $\omega(e)^{IJ}_{\mu}$ of General relativity to a larger group $G\sim SU(N)$, which breaks to $SU(2)\times SU(N-2)$ where the $SU(N-2)$ factor is identified with the dark sector \cite{unif}}. In this work we take a phenomenological perspective and simply assume that this dark sector exists and focus on the cosmological consequences.\\

In this model, late time acceleration arises from extending the color sector of QCD to have an ``invisible-copy'', namely $G_{D}=SU(3)_{D}$.  IQCD has similar quantum field theoretic properties of QCD, in that it is confining in the IR.  We will consider two species of quarks, up and down quarks, for minimality and a dark chiral symmetry. It is well known that pions arise as Goldstone modes associated to Chiral Symmetry Breaking (CSB), and in turn the microphysical description of CSB, the Nambu-Jona-Lasinio mechanism \cite{Nambu:1961}, is a strongly coupled version of superconductivity induced by hard gluon exchange.  
A key feature of our DE model uses the same physics of CSB in the Invisible QCD sector. During the matter and radiation era, the dark pions and gluons have negligible effects.  However, at late times the interaction energy between the dark pions and gluons become more significant and asymptote a nearly constant energy density, mimicking an effective cosmological constant.  Through the consistency of the coupled field equations, this interaction energy naturally leads to late time acceleration and we find an interesting connection between the scale of CSB and the scale of DE.       

In this work we find that IQCD in the context of a dark Big Bang Nucleosyntheiss context naturally predicts a dark matter candidate comprising of dark compact objects, such as an exotica of dark stars.  Intriguingly, the same meV scale of CSB that dictates late time acceleration is also the scale of hadronization of dark quarks.  These quarks hadronize eventually forming dark nuclei around 1Gyr during the epoch of galaxy formation.  We perform estimates that show the inevitable formation of exotic dark stars around the (visible) galaxy formation epoch. 
The precise structure formation in the dark sector deserves future numerical investigations beyond the scope of this paper
 but we will provide estimations of the essential astrophysics. We will discuss the observables our model in light gravitational microlensing constraints.  Finally, we will discuss phenomenological implications of our model, such as more accurate microlensing experiment, and by-fly anomalies in our solar system.  
 
\section{The Theory}

\noindent 
The $SU(N)_D$ gauge theory we are moving from has a behavior similar to an invisible copy of QCD, hence the name. IQCD assumes that in the dark sector chiral symmetry is broken at some scale $f_{_D}$ corresponding to the mass of a dark pion (dpion) $\pi_{_D}$. The gauge coupling becomes strong at the IR limit, triggering confinement and chiral symmetry breaking at a scale $\Lambda_D$. Below $\Lambda_D$, the effective theory is described by ``pions'' which are pseudo-Nambu--Goldstone bosons associated with the spontaneously broken global flavor symmetry of the hidden sector. At longer wavelengths, confinement still takes place, but we expect that atoms of DM start forming together with several other hadronic bound states, which stand as condensed phases of the invisible sector constituents. 

Labeling the indices of the adjoint representation of SU(3) as $A,B=1,...8$, the gauge field strength $F^A$ casts  
\be\label{F-general}
F^A_{~\mu\nu}=\partial_\mu A^A_{~\nu}-\partial_\nu
A^A_{~\mu}-g f^A_{~BC}A^B_{~\mu}A^C_{~\nu}, 
\ee
where $f_{IJK}$ are the structure constants of the $SU(3)$ algebra.
We are led to consider the most general gauge invariant action coupled to ``dark quarks'', the mass of which $m_q$ is a free parameter in our theory:
\bea 
&&\!\!\!\cal{L}\rm_{IQCD}\equiv\cal{L}\rm_{A} +\cal{L}\rm_{D} \nonumber\\
&&\!\!\!=-\frac{1}{4 g^2}F^A_{~\mu\nu}F_A^{~\mu\nu} + \sum_q \imath\,\overline{\psi}_q\, D_{\mu}\gamma^{\mu}\psi_q  - m_q \overline{\psi}_q  \psi_q \,. 
\eea
Here the sum is over the quark families labelled by $q$, $D_\mu$ stands for the covariant derivative with respect to the gravitational connection, $\gamma^\mu\!=\!\gamma^I\, e_I^\mu$ and the metric field is decomposed in tetrad, namely $g_{\mu\nu}=e^I_{\mu} \, e^I_{\nu}$, the inverse of which is denoted as $e^\mu_I$ and the internal $SO(3,1)$ indices of which are $I=1,2...4$. Given that we are working in a system where the dpion forms as a result of CSB, the decay constant $f_{_D}$ is defined through the coupling of the axial current to the dpion. In particular, dpions can be created by the axial isospin currents. \\

Below we summarize previous results on the homogenous cosmological dynamics of IQCD, showing a fixed point late-time acceleration of the Universe. Then we analyze the behavior of dark quarks and gluons on astrophysical scales, arguing that the invisible sector can also provide an interesting phenomenologically viable dark matter candidates. 
 

\subsection{Long wavelength modes and Dark Energy}

\noindent
The long wavelength modes behavior of the theory can be reckoned without loss of generality focusing on its subgroup, $SU(2)_{L}\times SU(2)_{R} \rightarrow SU(2)_{V}$ with its gauge field $A^a_{~\mu}$, where
$a,b=1,2,3$ and $\mu,\nu=0,1,2,3$ denote the dark color and space-time indices, respectively. The gauge field strength $F$ is
\be\label{F-general}
F^a_{~\mu\nu}=\partial_\mu A^a_{~\nu}-\partial_\nu
A^a_{~\mu}-g\epsilon^a_{~bc}A^b_{~\mu}A^c_{~\nu}, 
\ee
where $\epsilon_{abc}$ is the totally antisymmetric Levi-Civita symbol, the structure constant of the $SU(2)$ algebra.

Matrix elements of $\mathcal{J}_{\, a}^{5 \, I}(x)$ between the vacuum and an on-shell dpion states can be parametrized in terms of the dpion field as
\bea \label{expJ}
\langle  0| \mathcal{J}_{ \,a }^{5\, I }(x) | \pi^b \rangle=f_{_D}\, \delta^b_a \,\partial^I \pi_b(x) \,.
\eea
Axial vectors are space-like. We can rotate the expectation value of  $\langle  0| \mathcal{J}_{ \,a }^{5\, I }(x) | \pi^b \rangle$ within the internal Lorentz indices' space, so as to accomplish an explicitly space-like axial vector with vanishing temporal component \footnote{Models of dark energy with fermion axial condensates have been considered in \cite{Alex1,Alex2}}. The symmetry of the vacuum state on the FLRW background allows us to further reduce \eqref{expJ} to a homogenous axial vector:
\bea \label{expJ2}
\langle  0| \mathcal{J}_{ \,a }^{5\, I }(x) | \pi^b \rangle=f^2_{_D}\, 
\delta_{a}^i \,\pi(t, 0) \,,
\eea
where $\pi(t)\!\equiv\! || \pi^a(t)||$, with respect to the internal indices. The interaction of the axial current with the gauge field ${\cal L}^{ int.}_\pi = g \,A^{a}_{\mu}\,\mathcal{J}_{a}^{5\, \mu}$ is therefore consistent with homogenous and isotropic metrics.

The low energy dpion effective Lagrangian reads
\bea \label{pio}
\cal{L}\rm_{\pi}^0 = -\frac{1}{2} \partial_{\mu}\pi_a\partial^{\mu}\pi^a +\frac{\lambda}{4}\left(\pi^{a}\,\pi_{a}  - f^2_{D}\right)^{2}\,.
\eea
Consequently, the total effective Lagrangian reads
\begin{eqnarray} \label{lato}
{\cal L}_{ tot.} \!\!\!\!&=& \!\! {\cal L}_{\rm GR} +{\cal L}_{\rm A}+{\cal L}^{ int.}_\pi +{\cal L}^0_{\pi} \\
&=& \!\!\! M^2_p \,R \rm-\frac{1}{4} F^a_{~\mu\nu}F_a^{~\mu\nu} \! + g \,A^{a}_{\mu}\,\mathcal{J}_{a}^{5\, \mu} + \cal{L}\rm_{\pi}^0, \nonumber
\end{eqnarray}
in which we have introduced the reduced Planck mass as $M_p^2= (8 \pi G)^{-1}$. Quark fields have been integrated out in the path integral in order to get the effective action. The interaction term ${\cal L}^{ int.}_\pi = g \,A^{a}_{\mu}\,\mathcal{J}_{a}^{5\, \mu}$, which entails parity violations of the $SU$(2) subgroup of the dark sector, preserves renormalizability. The total action is ${\cal S}_{tot.}= \int d^4x \,e\,\, {\cal L}_{tot.}$, with $e\!=\!\sqrt{-g}$ denoting the determinant of the tetrad $e_\mu^I$.

\noindent Solutions to the field equations that are consistent with a FLRW background are recovered imposing a rotationally invariant configuration for the gauge field,  $A^a_{~\mu}= a(t)\,\phi(t)\,\delta^a_\mu\,$. Together with \eqref{expJ}, the ansatz allows us to recover the energy-momentum tensor of the theory, which is isotropic and homogenous, and yields the barotropic index $w=-1$. Indeed, energy and pressure densities respectively read 
\begin{eqnarray}
\rho_{_{\bf AJ}}\!\!&=& 3\,  g\, f_D^2 \, \phi(t) \,\pi(t), \nonumber \\
-\, P_{_{\bf AJ}} \!\!&=&3 \,g \, f_D^2\, \phi(t) \, \pi(t), \nonumber
\end{eqnarray} 
$g$ denoting above the absolute value of the coupling constant. Furthermore, the ansatz provides that the total gauge Lagrangian becomes 
\be
\!\!\mathcal{L}^{\bf A}+\, \mathcal{L}^{\rm int.}\!= \frac{1}{2\, g^2}\left( 3 \,(\dot{\phi}+H\phi)^2 -3 \, g^2 \phi^4\right)+ 3 g \phi\, \bar{J}(a), 
\ee
where $\bar{J}(a)\!\equiv\! f^2_{_D} \pi(t)$. It follows that the equation of motion for $\phi$, which now captures the dynamics of $A^a_{~\mu}$, can be cast as 
\be \label{eom1}
\ddot{\phi} + 3 H \dot{\phi}+ (2 H^2 +\dot{H}) \, \phi+ 2 \, g^{2} \phi^{3}  - g \, \bar{J}(a) =0 \,.
\ee

The equation of motion for the dpion field is recovered varying (\ref{pio}) within the assumption of spatial homogeneity. This is plausible, since a previous inflationary epoch of the universe can smooth out the dpion field. In the next section, we show that the dpion field remains homogenous against perturbations. 

Using the decomposition in a homogeneous absolute value (in the internal space) times a space-dependent unit vector, {\it i.e.} $\pi_a=||\pi_a|| \,n_a=\pi(t)\, n_a(x)$, we recover for the pion field
\be \label{eom2}
\ddot{\pi} + 3H\dot{\pi} + \lambda\, \pi(\pi^{2} -f^2_{D}) - 3 \,g f_D^2 \phi  
=0\,.
\ee
To gain some insight as to why we might expect to see late time acceleration, consider the slow roll regime of the dpion field, which is obtained by neglecting the acceleration term.  In this approximation, when the dpion field exhibits an inverse scaling with time, $\pi = \pi_0\, a^{-1}(t)$, the equation of motion reduces to
\be
3H\, \frac{\dot{a}}{a^2} = \frac{ \lambda }{a} \left( \frac{\pi_0^2}{a^2}- f^2_D \right)\,.
\ee
Solving this latter results in a power law acceleration of the Universe, namely $H(t)\simeq t^{-1}$, provided that $\pi(t_0)\!=\!\pi_0\!>\!\!>\!f_{_D}$ and, as customary when taking into account cosmological scalar fields, the slow roll condition holds: $3H\dot{\pi}\simeq V'\!>\!\!>\! \ddot{\pi}$. When the interaction term $\cal{L}\rm_{\rm int} \!=\! g\,\phi(t)\, f^2_{_D} \pi(t)$ between the dpion and the gauge field is considered, we will see that this term persists to have a nearly constant energy density yielding a negative pressure equation of state. Finally, it is straightforward to show that a slightly different behavior in the time dependence of the dpion, {\it i.e.} $\pi \!=\! \pi_0 \,a^{-n}(t)$ with $n>0$, would yield the same late time-behavior $H(t)\!=\!t^{-1}$.\\  

Late time acceleration is recovered when the gauge field asymptotically evolves in time as the scale factor and the pion field approach the constant value $\pi\simeq f_D$. 

Under customary assumption, we are able to solve for the coupled system of differential equations in the configuration space $\{\phi(t), \pi(t) \}$, and to find solutions consistent with a de Sitter expanding phase.

Both solutions for $\phi$ and $\pi$ monotonically decrease and converge asymptotically towards values that are proportional to $f_D$; thus their product conspires to provide an accelerating solution well approximated by a de Sitter phase, the effective cosmological constant of which assumes the asymptotic value
\be
\Lambda \simeq \frac{f_D^4}{M_p^2} \,.
\ee
Supernovae data, which entail at current times $H\simeq 10^{-42}$~GeV, are consistent with the asymptotic value for the gauge field $A \!\simeq\! f_D\!\simeq\!~\!10^{-3}\!$ eV, the coupling constant $g$ is assumed to be order unity. This suggests a fascinating conclusion: cosmic acceleration is the result of CSB in the dark sector, since it occurs at the same scale of energy, {\it i.e.} $M_{_{DE}}\simeq f_D$. 

Consistency of the solutions can be checked: conservation of the energy-momentum tensor is achieved, and a cogent analysis of the attractor behavior of the solutions further corroborates these results. For this purpose, we focus on the equations of motion for $\phi$ and $\pi$, and consider a de Sitter phase of expansion of the Universe (namely the value of the Hubble parameter is constant, and actually $H=10^{-33}eV$). We derive for $g\!=\!\lambda\!=\!0.1$ the numerical value 
\begin{equation}
(\phi_f,\pi_f, H_f)\!=\!(2.17\cdot 10^{-3} {\rm eV}, \, 2.04\cdot 10^{-3} {\rm eV},1.08 \cdot10^{-35} eV). \nonumber
\end{equation}
from the relations at the fixed point:
\begin{equation}
2g^2 \phi_f^3-gf_D^2\pi_f=0,
\end{equation}
\begin{equation}
\lambda \pi_f(\pi_f^2-f_D^2)-3gf_D^2\phi_f=0,
\end{equation}
and 
\begin{equation}
H_f^2M_p^2+\frac{1}{2}H_f^2\phi_f^2+\frac{1}{2}g^2\phi_f^4-gf_D^2\phi_f\pi_f-\frac{\lambda}{12}(\pi^2-f_D^2)^2=0\,.
\end{equation}
The stability of the fixed point can be then analyzed in the conventional way introducing the variable $N=\ln a(t)$. The system of differential equations, taking into account respectively the equations of motion for the gauge field $\phi$, the pion field $\pi$ and the evolution of the Hubble parameter with the second Friedmann equation, then reads 
\begin{equation}\label{3}
H^2\frac{d^2\phi}{dN^2}+3H^2\frac{d\phi}{dN}+2H^2\phi+2g^2 \phi^3-gf_D^2\pi=0,
\end{equation}
\begin{equation}\label{4}
H^2\frac{d^2\pi}{dN^2}+3H^2\frac{d\pi}{dN}+\lambda \pi(\pi^2-f_D^2)-3gf_D^2\phi=0,
\end{equation}
and 
\begin{eqnarray}\label{5}
&H\frac{dH}{dN}=-H^2M_p^2-\frac{1}{2}(H\frac{d\phi}{dN}+H\phi)^2-\frac{1}{2}g^2\phi^4+gf_D^2\phi\pi\nonumber\\
&-\frac{1}{3}(H\frac{d\pi}{dN})^2+\frac{\lambda}{12}(\pi^2-f_D^2)^2.
\end{eqnarray}
We denote the derivatives of the fields as $\varphi\equiv\frac{d\phi}{dN}$ and $\psi\equiv\frac{d\pi}{dN}$, and  consider the perturbation $\phi=\phi_f+\delta \phi$ for the electromagnetic field and $\delta \varphi= \frac{d\delta \phi}{dN}$ for its derivative, and the perturbation $\pi=\pi_f+\delta \pi$ for the pion field and $\delta\psi=\frac{d\delta\pi}{dN}$ for its derivative, and the perturbation for the Hubble parameter $\delta H$. Relations (\ref{3}-\ref{5}) reduce to
\begin{equation}
\frac{d}{dN}\left( \begin{array}{c}
\delta \phi \\
\delta \varphi\\
\delta \pi\\
\delta \psi\\
\delta H
\end{array} \right)=
\mathcal{M}
\left( \begin{array}{c}
\delta \phi \\
\delta \varphi\\
\delta \pi\\
\delta \psi\\
\delta H
\end{array} \right)\,,
\end{equation}
in which
\begin{widetext}
\begin{centering}  
\begin{equation}
\mathcal{M}=
\left( \begin{array}{ccccc}
0 & 1 & 0 & 0 & 0 \\
  -2-6g^2\frac{\phi_f^2}{H_f^2}+\frac{\phi_f^2}{M_p^2}+2g^2\frac{\phi_f^4}{H_f^2M_p^2}-g\frac{f_D^2\phi_f\pi_f}{H_f^2M_p^2} & -3+\frac{\phi_f^2}{M_p^2} & g\frac{f_D^2}{H_f^2}-g\frac{f_D^2\phi_f^2}{H_f^2M_p^2}-\frac{\lambda \phi_f \pi_f (\pi_f^2-f_D^2)}{3H_f^2M_p^2} & 0 &\frac{\phi_f^3}{H_f M_p^2}\\
  0 & 0 & 0 & 1 & 0 \\
  3g\frac{f_D^2}{H_f^2} & 0 & -3\lambda\frac{\pi_f^2}{H_f^2}+\lambda\frac{f_D^2}{H_f^2} & -3 & 0 \\
  -\frac{H_f\phi_f}{M_p^2}-2g^2\frac{\phi_f^3}{H_f M_p^2}+g\frac{f_D^2\pi_f}{H_f M_p^2} & -\frac{H_f\phi_f}{M_p^2} & g\frac{f_D^2\phi_f}{H_f M_p^2}+\frac{\lambda  \pi_f (\pi_f^2-f_D^2)}{3H_f M_p^2} & 0 & -2-\frac{\phi_f^2}{M_p^2}
\end{array} \right)\,.
\end{equation}
\end{centering}  
\end{widetext}
The eigenvalues of $\mathcal{M}$ are easily found to be $\lambda_1=-1.5+\imath 1.01228\times10^{32}$,$\lambda_2=-1.5-\imath 1.01228\times10^{32}$, $ \lambda_3=-1.5+\imath4.63919\times10^{31}$, $ \lambda_4=-1.5-\imath 4.63919\times10^{31}$ and $\lambda_5=-2$. Since the real parts of the eigenvalues of $\mathcal{M}$ are all negative, the fixed point is shown to be stable. 

In order to corroborate the analysis reported above, we show in Fig.~1 results from numerical integrations of the dynamical system, which are fully consistent with previous claims.

\begin{figure}[htb] \label{figuna}
\begin{center}
\caption{Plot of $w$ against the redshift $z$, from current time up to recombination. The coupled system of non-linear differential equations has been solved numerically for $g\!= \!10^{-1}$ and $\lambda\!=\!1$, and the initial conditions on the functions $H(0)\!=\! 10^{-42} GeV\,$ and  $\phi(0)\!\simeq\!\pi(0)\!\simeq\!f_D \!=\!10^{-3} \, eV\,$, and on their derivatives $\phi'(0)\!\simeq\! \pi'(0)\!=\! 10^{-5}\, (eV)^2\,$. Transition from DE to (dark sector) radiation happens for $z\simeq2$ (blue line). With a choice of the coupling constant $g$ an order of magnitude smaller, transition to DE happens at $z\simeq9$ (red line).}
\includegraphics[scale=0.72]{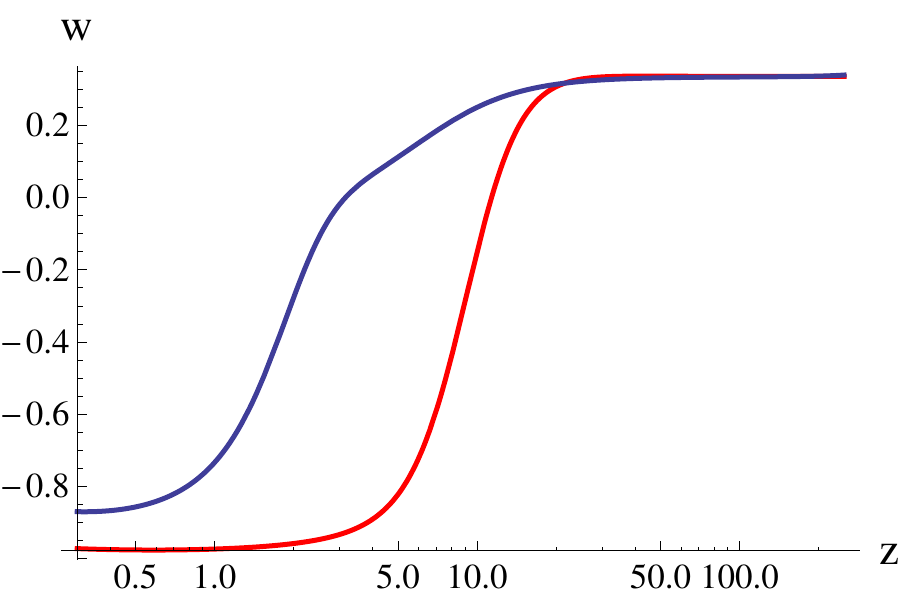}  
\end{center}
\vspace{-4mm}
\end{figure}

\subsection{Short wavelength modes and Dark Matter}

\noindent 
We now analyze shorter wavelength modes of the dark sector to seek a viable Dark Matter scenario. 
In the dark sector, the chiral symmetry is dynamically broken by non-perturbative effects of dark strong interactions, such that  $f_{D}$ is close to the dark QCD scale $\Lambda_{D}\simeq f_{D}\simeq 10^{-3}$. As a consequence, the dark quarks $\tilde{q}$ will hadronize into dark mesons and baryons with an effective scale $L\simeq \Lambda^{-1}_{D}\simeq 1\, {\rm mm}$ and are assumed to be a singlet with respect to the standard model gauge group\footnote{ The dark quarks transform in the $(3,1,1,1)$ representation of $G=SU(3)_{D}\times SU(3)_{c}\times SU(2)_{L}\times U(1)_{Y}$}. It is the physics of these dark quarks and gluons that will pave the way to a novel perspective on DM and potential new observational windows for detecting them.

Let us introduce for minimality only two dark quarks $\tilde{u},\tilde{d}$, taken in the fundamental representation of $SU(3)_{D}$, and consider the following extension to the Lagrangian:
\begin{equation}
\mathcal{L}_{\tilde{q}}=m_{\tilde{u}}\bar{\tilde{u}}\tilde{u}+m_{\tilde{d}}\bar{\tilde{d}}\tilde{d}+h.c\,.
\end{equation}
We assume that dark quarks only interact strongly with each other, {\it i.e} no ordinary electromagnetic and weak channels are present in the hidden sector. 

Because the dark electric charge is absent, dark neutrons and protons are degenerate bound states with an effective radius $1\, {\rm mm}$.  Furthermore, Big Bang Nucleosynthesis (BBN) in the dark sector is simpler than the visible sector. 
In the visible sector, BBN is crucially connected to the proton to neutron ratio, which is generated by weak interaction processes after freeze-out. However, in our scenario there is no dark $\gamma,W,Z$ and $BBN$ is catalyzed solely by strong nuclear interactions. Consequently, the reaction chain will simply produce the most stable dark nuclei with practically $100\,\%$ abundance.  Also, the possibility to produce higher atomic mass nuclei cannot be neglected \footnote{For minimality, dark quark masses can be introduced by hand as free parameters without any inconsistency with renormalization. On the other hand, the dark IQCD can be thought as a residual gauge symmetry of a higher gauge group. In this case, quark masses can be generated by Higgs mechanisms. }.

Similar to the visible sector, the highest binding energy is for He-4, composed of two neutrons and two protons in the lowest nuclear orbital,$1s-1s$. This configuration minimizes the angular momentum and energy, a well known fact in nuclear physics. So the dark BBN processes will proceed as follow: 
\begin{enumerate}
\item
Dark hadronization at a temperature of $10^{-3}\div10^{-4}eV$.

\item
Formation of dark deuterium $\widetilde{D}$ from proton and neutron collisions.
\item
Formation of dark $\widetilde{He}$-4 from the sequence of reactions:
\begin{eqnarray}
&(\widetilde{p},\widetilde{n})+\widetilde{D}\rightarrow (\widetilde{He}^{3},\widetilde{H}^{3}),\nonumber\\
&(\widetilde{p},\widetilde{n})+(\widetilde{He}^{3},\widetilde{H}^{3})\rightarrow \widetilde{He}^{4}\,. \nonumber
\end{eqnarray}

\end{enumerate}

Dark hadronization occurs around a redshift of $z\simeq 1$ for $\Lambda_{D}\simeq 10^{-3}\, \rm eV$, corresponding to $1\, \rm Gyrs$. In other words, dark quarks hadronize around the epoch of galaxy formation.  Compared to WIMP candidates, where the correct abundance ariises {\it \`a la} the WIMP miracle, in this scenario dark quark masses determines the DM mass density $\rho_{DM}$ \footnote{In fact, after the inflationary reheating epoch, the number densities of dark and visible particles  are expected to be the same, {\it i.e} we assume that inflaton decays equally reheat both the dark and visible sector. However, $m_{\widetilde{He}}\simeq 5\, \rm GeV$ automatically guarantees a current abundance of Cold Dark Matter from Dark BBN if and only if the inflaton field will democratically reheat both the ordinary and the dark sectors. In general this hypothesis can be relaxed. An asymmetric reheating mechanism between the ordinary sector and IQCD sector can establish different initial reheating temperatures among the two sectors. In fact, the inflaton field can be coupled differently with ordinary SM particles and dark quarks. In this case, the cosmological scenario can be more complicated. Generically, for $\tilde{T}_{R}<T_{R}$, $m_{\tilde{He}}$ is smaller than 5GeV. A different initial reheating mechanism can also significantly influence the dark BBN history. For DM phenomenology, we cannot consider dark quarks
lighter than $1\,\rm keV$ (warm dark matter). These complications are expected to not relevantly change main results of this section. A more complete analysis in full generalities deserve future investigations beyond the purposes of this letter. }.  For a cold thermal dark halo of velocity $v_{0}\simeq 220\, {\rm km/s}$ the CM energy of $\tilde{He}-\tilde{He}$ are $E_{CM}\sim 1\, \rm keV$ which is above the scale of dark hadronization, $\Lambda_{D}$ \footnote{As a consequence, $\Omega_{DM}/\Omega_{b}\simeq 5$ can be accounted for if $m_{\tilde{He}}\simeq 5\, \rm GeV$. The small scale of chiral symmetry breaking $\Lambda_{D}\simeq 10^{-3}\div10^{-4}\, {\rm eV}$, will give a negligible contribution to nuclei masses, which implies $m_{\tilde{He}}\simeq 2m_{p}+2m_{n} \simeq 6 m_{u}+6m_{d}$, {\it i.e} $m_{\tilde{u}}\simeq m_{\tilde{d}}\simeq 0.3 \, \rm GeV$.}.  Thus, dark quarks are practically unbounded in thermal collisions and dark QCD is in the perturbative regime where strong interactions are dominated by quark-gluon exchange.  

 As the universe continues to cool a thermal halo of dark Helium could form.  However such a halo is excluded by stringent limits on DM self-interaction cross section. Direct constraints from galaxy clusters 1E 0657-56 (Bullet Cluster) put limits of $(\sigma/m)_{s.c}<1g/cm^{3}$ , where $\sigma$ is the dark particle self-cross section, and $m$ is the dark particle mass\cite{Harvey:2015hha,Markevitch:2003at}. This cross section is much smaller than a typical QCD cross section, which is typically larger than $\sigma_{QCD}/m \gg 10^{14}g/cm^{3}$ \footnote{Recall that a Rutherford electromagnetic scattering of two protons correspond to $\sigma_{e.m}/m\sim 10^{14}g/cm^{3}$ and larger for QCD.} .\\

However, this scenario can be reconciled with the Bullet cluster as follows: If the dark halo of the Bullet cluster is mostly recombined into compact objects, like dark stars, the collision probability will be less than for WIMPs. But, these compact objects would be detected by gravitational lensing measurements of the galactic dark halo. Limits recovered by MACHO and EROS collaborations exclude the prevalence of Massive Compact Halo Objects in our dark halo, with $M_{MACHO}>10^{-7}\, {\rm M_{\odot}}$ \cite{Alcock:2000ph, Tisserand:2006zx}. 

In our scenario, it is important to establish that the formation of exotic dark stars is rapid compared to standard structure formation.   For temperatures higher than the dark confinement scale, dark quarks and gluons are in a dark quark-gluon plasma phase, rapidly cooling due to thermal Bremsstrahlung. The energy loss for the thermal Bremsstrahlung contribution is  $dE_{Bremsstrahlung}/dt dV\sim n^{2}T^{1/2}$ where $n$ is the quark density while $T$ is the temperature of the dark thermal bath.  For $T>\Lambda_{D}$, this contribution is dominant to the recombination loss $dE_{rec}/dt dV\sim n^{2}T^{-1/2}$. For a system with a temperature $T\simeq GMm_{\tilde{q}}/R$ much higher than $\Lambda_{D}$, we can estimate the cooling time as 
\begin{equation}
t_{cool}\simeq \frac{nT}{dE/dt dV}= \frac{\alpha_{D}^{3}}{n}T^{1/2}\Lambda_{D}^{3/2}
\end{equation}
while the time scale for gravitational collapse is 
\begin{equation}
t_{grav}\simeq \left( \frac{R^{3}}{GM} \right)^{1/2}
\end{equation}
For an efficient cooling, $t_{cool}<t_{grav}$ leading to the constraints
\begin{equation}
R<\bar{R} \simeq O(1\div 10^{-3})\,\zeta^{-3/2} \bar{R}_{o}
\end{equation}
\begin{equation}
M>\bar{M}\simeq O(1\div 10^{-3})\,\zeta^{1/2} \bar{M}_{o}
\end{equation}
where $\zeta=\Lambda_{D}/m_{e}$, $m_{e}$ is the electron mass,
 $\bar{R}_{o}\simeq 74\, \rm kpc$, $\bar{M}_{0}\simeq 10^{44}\, \rm gm$
 are critical mass and radius of standard structure formation, defined as 
 \begin{equation}
\bar{R}_{o}\simeq \alpha_{e.m} \alpha_{G}^{-1}\lambda_{e}\left( \frac{m_{p}}{m_{e}}\right)^{1/2}
\end{equation}
\begin{equation}
\bar{M}_{o}\simeq \alpha_{e.m}^{5} \alpha_{G}^{-2}m_{p}\left( \frac{m_{p}}{m_{e}} \right)^{1/2}
\end{equation}
$\bar{R}_{o},\bar{M}_{o}$ are controlled by electromagnetic interactions among protons and electrons, 
while for $\bar{R},\bar{M}$ by dark strong confinement scale, dark strong coupling
and dark quark mass. 
The density needed for ordinary and dark structure formation is almost the same $\langle \rho \rangle \simeq 10^{-25}\, \rm gm\, \rm cm^{-3}$.
Assuming the structures are formed with overdensities of the order $\rho_{over}\simeq 100 \times \langle \rho \rangle$, the dark galaxy formation will have to be happen where the dark density was $1000$ times the density of present epoch. 
In the dark sector, $\bar{R}>r_{Hubble}$ while the mass needed is approximately $10^{40}\, \rm g$. 
As a consequence, dark structure formation begins quickly after the 
Invisible QCD  phase transition, so in redshift $z\simeq 1\div 10$, depending on the initial conditions (see Fig.1). 
a system with $3\times 10^{40}\, \rm gm$ rapidly cools and forms gravitational structures as a fragmentation process. 

In the following discussion, we will estimate the mass of dark compact objects in our scenario.  As mentioned above, the dark strong length scale is $l\simeq 1\, {\rm mm}$. Nuclear reactions can occur through quantum mechanical tunneling when the De Broglie wavelength $\lambda_{deB}\simeq l\times (c/v)$ overlap, where $v$ is velocity of colliding particles. 

After dark BBN the typical velocities in dark primordial thermal bath is $v\simeq 10\div 100\, \rm km/s$, $\lambda_{deB}\simeq 10^{3}\div 10^{4} l\simeq 1\div 10\, \rm m$. As a consequence, nuclear reactions will need an initial momentum of the order of $p_{Nucl}^{DM}\simeq 10^{-3}\div 10^{-4}\Lambda_{D}\simeq 10^{-7}\div 10^{-8}\, \rm eV$. The gravitational energy for particle needed is $$\epsilon_{g}^{DM}\simeq E_{grav}/\mathcal{N}_{\tilde{N}}\simeq Gm_{\tilde{N}}^{2}\mathcal{N}_{\tilde{N}}/R^{DM} \simeq \epsilon_{Nucl}^{DM}$$ where $\mathcal{N}_{\tilde{N}}$ is the number of dark nucleons in the star and $R^{DM}$ is the star radius. We can conveniently rewrite the expression  as 
$$\epsilon_{g}^{DM}\simeq \left(\frac{4\pi}{3} \right)^{1/3}Gm_{\tilde{N}}^{2}\mathcal{N}_{\tilde{N}}^{2/3}n_{\tilde{N}}^{1/3}$$
where $n_{\tilde{N}}=3\mathcal{N}_{\tilde{N}}/4\pi R^{2}$ is the number density. 

However, $\epsilon_{g}^{DM}/\epsilon_{g}^{ordinary}\simeq \epsilon_{nucl}^{DM}/\epsilon_{nucl}^{ordinary}\simeq 10^{-7}\div 10^{-8}eV/1KeV$.  Assuming the same density of nucleons, $n_{\tilde{N}}\simeq n_{N}$, one sees that a dark star is formed with a much smaller number of dark nucleons relative to ordinary stars, $\mathcal{N}_{\tilde{N}}\simeq 10^{-13}\div 10^{-19}\mathcal{N}_{N}$; where for ordinary stars, a number of $\mathcal{N}_{N}$ in a certain radius $R$ is usually needed for star formation.  Furthermore, a star's radius and mass rescales with the number of particles as, $ R^{DM}/R^{ordinary}\simeq [\mathcal{N}_{\tilde{N}}/\mathcal{N}_{N}]^{1/3} \simeq 10^{-4}\div 10^{-5}$
 and $ M^{DM}/M^{ordinary}\simeq [\mathcal{N}_{\tilde{N}}m_{\tilde{N}}^{2}/\mathcal{N}_{N}m_{N}^{2}]\simeq 10^{-14}\div 10^{-15}$. Therefore, dark stars are endowed with small masses and radii compared to ordinary stars \footnote{ Let us remark that there no dark photons in our scenario, so that dark stars have not a dark electromagnetic luminosity and electromagnetic scatterings not transit core heat to external dark star layers.  So that, heat can be only transmitted by strong interactions. For instance a reheated dark nucleon can exchange dark gluons within a radius separation from them $1\, \rm mm$. We expect that this dramatically changes the structure of dark stars with respect to ordinary stars, but this technical aspect is beyond the purpose of this paper. }.

 One could further expect that a dark star life-time to be much smaller than ordinary ones. We can roughly estimate their lifetime of the nuclear burning phase as $t_{star}^{DM}\simeq \epsilon M^{DM}/\mathcal{L}$, where $\mathcal{L}$ is the intensity factor 
parametrizing the propagation of heat in the dark star, while $\epsilon\sim 1\%$ is the fraction of the rest-mass energy  of He-4 available for nuclear reactions. To calculate $\mathcal{L}$ near the core is highly non-trivial. However, at longer radii, the intensity factor can be estimate as $\mathcal{L}\simeq R^{4}T^{4}/(\langle\sigma \rangle N)\simeq T^{4}R l_{col}$, where $\langle \sigma \rangle$ is the average opacity in the dark QCD plasma felt by gluons, $l_{col}\sim l$ is the average collision path roughly equal to the confinement scale, $T$ is the temperature of the dark star.   As a result the estimation of the lifetime,$t_{star}^{DM}$, will be proportional to $\Lambda_{DM}$, indirectly entering in all quantities involved such as the radius, temperature, luminosity and cross section.  Therefore, $t_{star}^{DM}$ is expected to be smaller than ordinary stars by a factor $10^{-12}$, {\it i.e} around $10^{-6}\, {\rm yrs}$. Let us keep in mind that this is a rough order of magnitude estimate.

An evaluation of the dark star lifetime is strongly dependent on the kind of star and its mass. These aspects deserve future numerical simulations. However, we expect that after a Gyr, practically all initial stars will be converted into stellar remnants. In particular, gravitational pressure cannot be balanced as usual by electronic Fermi pressures as in dwarfs, simply because there are not dark electrons. On the other hand, these objects are far from the Schwarzschild limit, so that black hole formation from dark supernovae is expected to be impossible. The only counter-pressure to gravity can come from nucleon or quark Fermi pressure. In other words, we expect that practically all dark stars will be converted to dark neutron stars of dark quark/strange stars. Eventually, black holes can be formed by the merging of two dark neutron stars or dark quark/strange stars, especially in dark binary systems. 

 Dark stars have an average mass of $10^{-14}\div 10^{-15}\,M_{\odot}\simeq 10^{16}\div 10^{17} {\rm Kg}$ This scale completely avoids any constraints by MACHO and EROS collaborations. A halo with so high a frequency of supernovae will be dominated by dark neutron stars, dark strange/quark stars and black holes. Peculiarly, they will be very compact objects, the radii of which will be a factor $10^{-4}$ smaller than ordinary radii, {\it i.e} $r_{MACHO}\simeq 10^{2}\div 10^{3}\, {\rm km}$. As a consequence, dark halos are expected to be mainly composed of dark neutron stars with a self-collision probability $\sigma/m\simeq \pi (2r_{MACHO})^{2}/M_{MACHO}\sim 10^{-15}\div 10^{-16}\, {\rm cm}^{2}/{\rm g}$, estimated as a disk scattering of radius $r_{MACHO}$. Thus, IQCD can also provide a good candidate for DM, composed of very compact heavy objects rather than WIMPs.
 
Such a rapid star formation will change the dark galaxy dynamics with respect to ordinary structure formation. In standard theory, star formation is slower than formation of the Silk disk, so that spiral galaxies are formed, while elliptic ones are probably formed by the merging of spiral ones \cite{Galaxy}. In our case, the situation is different. Dark star formation is so fast that a halo of dark stars is expected to be an elliptic one. In a broad sense, such compact stars can be thought of as composite WIMPZILLAs, forming a thermal dark halo with very small cross sections. This is a crucial propriety of our DM candidate, otherwise a dark disk cannot correctly account for the observation of velocity rotation curves. From velocity rotation curves of our Galaxy, one can estimate the number of these compact objects in our dark halo as $\bar{N}=M_{DM}/M_{DarkStar}\simeq 10^{12}M_{\odot}/(10^{-14}\div 10^{-15}M_{\odot})\sim 10^{26}\div 10^{27}$. Naively, one could  estimate the average interdistance of these objects as $d_{DarkStar}\simeq (R_{DH}/\bar{N}^{1/3})\sim 15\,{\rm kpc}/10^{8\div 9}\sim 10^{-4}\div 10^{-5} \,{\rm pc} \simeq 10^{9}\div 10^{8}\, {\rm km}$, where $R_{DH}$ is the radius of the dark halo. This estimation is a rough approximation, because we know that Dark halo mass is mostly concentrated in a radius of only $4\, \rm kpc$ ($90\%$ of the total mass or so). But it is enough to get an order of magnitude in order to compare to usual ordinary numbers.
   \\

Indirect hints of such a candidate could be the observation of objects formed much earlier than predictions from standard structure formation theory such as  supermassive black holes and early supernovae \cite{Dolgov:2014vba}. In fact, the first dark neutron stars can be formed just $10^{-6}\, {\rm yrs}$ after the Big Bang, so that they provide an exotic gravitational seed, favoring, for example, the formation of black holes at high redshift $z\sim 5\div 10$ (old as $0.4\div 1\, {\rm Gyr}$).

Our scenario also strongly motivates new experiments measuring masses of planets and sun in the solar system. For instance, high precision measures of masses can be done with GPS satellites, or by fly anomalies. Let us note that the presence of our dark objects in the solar system is not ruled-out only by current data. For instance, we know that the mass of the Earth itself is roughly $10^{-6}\, \rm M_{\odot}$, measured with a precision of one part of million, {\it i.e} $10^{-12}\, \rm M_{\odot}$ \cite{Earth}. The average mass of dark stars is $10^{-14}\div 10^{-15}\, \rm M_{\odot}$, as said before. In principle, dark stars can be captured in the gravitational potential of the solar system, and mainly by the Sun. Average thermal velocity of dark stars is expected to be $v_{0}\simeq 170 \div 270\, \rm km/s$ (similarly to WIMPZILLAs) \cite{Belli:2002yt}. Their distribution in our solar system can be highly non-trivial.  Hence, they are not easy to be captured by a Sun and planets after few gravitational scatterings. 

However, during several Gyrs, one could imagine that a subdominant ratio of dark neutron stars or quark/strange stars were so slowed by several gravitational scatterings to be captured by a Sun or planets.   Therefore, we suggest measuring their presence by looking for GPS anomalies of satellites in nearby planets in our solar system. For example, it is possible that planets and the Earth itself have captured dark stars as dark satellites. In particular, it is possible that a planet could have captured a dark ring of these objects, similar to the one around Saturn. Intriguingly, the formation of a ring with $10^{7}$ dark neutron stars or quark/strange stars around the Earth could be detectable in the next generation of experiments.  An estimate of the local DM density given by current constrains on galactic velocity curves is $\rho_{DM}\simeq 0.1 \div 1 \rm GeV/cm^{3}$ \cite{Belli:2002yt}, with central values $0.3\div 0.4 \rm GeV/cm^{3}$ for isothermal halo, corresponding to a number of dark neutron stars or dark quark/strange stars in our solar system of $10^{8}\div 10^{9}$. It is reasonable to assume that a ratio $1\div 10\%$ of them could be bounded in trapped orbits inside the solar system, i.e $10^{6}\div 10^{8}$ dark objects. Possible deviations of masses in our solar system with respect to expectations could provide a hint in favor of our model.

\section{Conclusion and Discussion} 

\noindent 
In this paper, we analyzed the implications of a dark QCD sector with a confinement scale $\Lambda_{D}\simeq 10^{-3}\div 10^{-4}\, \rm eV$. Continuing the analysis of the model recently proposed in \cite{IQCD}, we have demonstrated that DE can arise from the dynamical formation of dark strong condensates, due to the breaking of a dark chiral symmetry. On the other hand, the low confinement scale allows the formation of exotic dark stars with a mass of $10^{-14}\div 10^{-15}M_{\odot}$. These objects are so small to be eluded by present limits from gravitational microlensing and cluster limits. We are then tempted to dub these new objects Mini-MACHOs. 

We also demonstrated how the current abundance of DM can be easily recovered without the need of WIMP miracles or non-thermal mechanisms. This new paradigm for a unifying picture of DE and DM suggests several observables with respect to the standard picture. From the DE side, the dynamical evolution of the condensate energy density could be detected in future experiments testing late Universe expansion. On the other hand, Mini-MACHOs could be detected in more accurate microlensing experiments, or searching for by-fly anomalies in our solar system.  It is interesting that both DE and DM are governed by the same scale of Dark Chiral Symmetry Breaking.

Let us comment that our scenario is capable of predicting gravitational waves signals detectable by LIGO/VIRGO collaboration in the near future. In fact, even if the gravitational masses of dark exotic stars are so small, they are spread out in the dark halo with a much higher number and density then ordinary stars. In particular, such as rapid supernovae events in the dark sector can release characteristic signals with very different features compared to ordinary supernovae. On the other hand, the presence of a large number of Mini-MACHOs in the dark halo can increase the probability of dark binaries emitting gravitational waves. 

Of course, a further theoretical question worth addressing is: Why does the dark QCD sector contribute to the vacuum energy while standard QCD is somehow to be screened by an unknown mechanism?  A possible explanation could be related to an unknown UV completion of our model. Inspired by Ref.~\cite{Dvali:2007nm}, we are tempted to suggest that if the SM is localized on a Brane-world or a domain wall in higher dimensions while IQCD is localized on a Dark parallel D-Brane-World or Dark domain wall, the vacuum energy can be differently subtracted by different D-Brane/Wall tensions or different RR or NS-NS fluxes in the Calabi-Yau internal compactifications wrapped by branes.

{\it{\textbf{Acknowledgement.}}} 
We are indebted G.~Dvali and D.~Spergel and Ian Dell' Antonio for enlightening discussions.
A.~A. and A.~M. would also like to thank LMU for hospitality during the preparation of this paper. Work of A.~A. was supported in part by the MIUR research grant Theoretical Astroparticle Physics PRIN 2012CPPYP7 and by SdC Progetto speciale Multiasse La Societ\`a della Conoscenza in Abruzzo PO FSE Abruzzo 2007-2013. A.M. wishes to acknowledge support by the Shanghai Municipality, through the grant No. KBH1512299, and by Fudan University, through the grant No. JJH1512105.

\end{document}